\documentclass{ws-procs975x65}
\begin{document}

\title{BROKEN SCALE INVARIANCE AND QUINTESSENCE
\\
({\it a quarter of a century ago})}

\author{GIOVANNI VENTURI}

\address{Dipartimento di Fisica, Universit\`a degli Studi di
Bologna, and INFN, via Irnerio 46, 40126 Bologna, Italy.
\email{armitage@bo.infn.it}}
\begin{abstract}
The cosmological consequences of a simple scalar field model for the generation of Newton's constant through the spontaneous breaking of scale invariance in a curved space-time are again presented and discussed. Such a model leads to a consistent description wherein the introduction of matter introduces a small perturbation on a de Sitter universe and a time dependence of the gravitational coupling.
\end{abstract}

\bodymatter
\vspace{0,8 cm}
A quarter of a century ago we suggested a simple scalar field model for the generation of the gravitational coupling (Newton's constant) through 
the spontaneous breaking of scale invariance in a curved space time\cite{Cooper:1982du}. In this
brief note we wish to revisit the  
model which consisted of the following globally scale invariant Lagrangian density for a scalar filed $\sigma$
\begin{equation}
L=-{1\over 2}g^{\mu\nu}\partial_{\mu}\sigma\partial_{\nu}\sigma-{\lambda\over 4}\sigma^4+{\gamma\over  
2}\sigma^2R+L_m
\end{equation}
where $\gamma$, $\lambda$ are dimensionless positive constants ($\lambda>0$ for stability),  $L_m$ is the matter Lagrangian density (does not contain $\sigma$) and $R$ the curvature  
scalar. In flat space one can have a non-zero vacuum expectation value for
$\sigma$ ($\langle\sigma\rangle_0\neq 0$), in our   
case we assume the vacuum is a scalar particle condensate and examine the cosmological consequences.
We consider a Robertson-Walker line element
\begin{equation}
ds^2=-dt^2+a^2(t)\left ({dr^2\over 1-kr^2}+r^2d\theta^2+r^2\sin^2\theta d\phi^2\right)\end{equation}
and matter behaving like an isentropic perfect fluid having energy-mo\-men\-tum tensor
\begin{equation}
T^{\alpha\beta}_m=pg^{\alpha\beta}+(\rho+p) u^{\alpha} u^{\beta}
\end{equation}
where $\rho(t)$, $p(t)$ and $u^{\alpha}$ are the energy density, pressure, and velocity four-vector, respectively.  
The Einstein equations obtained from Eq. (1) are
\begin{equation}\label{E1}
\dot{\rho}=-3{\dot{a}\over a}(\rho+p)
\end{equation}
\begin{equation}\label{E2}
{\dot{a}^2\over a^2}+{k\over a^2}={\rho\over 3\gamma\sigma^2}+{1\over 6\gamma}{\dot{\sigma}^2\over  
\sigma^2}-2{\dot{a}\over a}{\dot{\sigma}\over \sigma} +{1\over 12\gamma} \lambda\sigma^2
\end{equation}
\begin{equation}\label{E3}
{d\over dt}(\sigma\dot{\sigma}a^3)={(\rho-3p)\over 6\gamma+1}a^3.
\end{equation}
where the dot denotes differentiation with respect to the time $t$.

In the absence of matter $(\rho=p=0)$ one obtains a time independent "vacuum" solution given by
\begin{equation}
\sigma=\sigma_0=\left ({\gamma R_0\over \lambda}\right )^{1/2}
\end{equation}
where $R_0$ is a constant and corresponds to a flat space ($k=0$) of constant curvature $R_0/12$ (de Sitter),
\begin{equation}
a=a_0(t)=a_0 (0) {\rm exp}(H_0 t)
\end{equation}
\begin{equation}
H_0=\left ({\lambda\over 12\gamma}\sigma_0^2\right )^{1/2}=\left ({R_0\over 12}\right)^{1/2}.
\end{equation}
Further, from the weak-field limit\cite{Turchetti:1981zb}, Newton's coupling constant $G$ is given by:
\begin{equation}
{1\over 8\pi G_0}=\gamma \sigma_0^2{6\gamma+1\over 8\gamma+1}
\end{equation}

One now treats the introduction of matter as a small perturbation and writes
\begin{equation}
\sigma=\sigma_0[1+\chi(t)],
\end{equation}
\begin{equation}
a=a_0(t)+\delta a(t),
\end{equation}
where $\chi$ and $\delta a$ are assumed small. We keep $k=0$ and assume an equation of state $p=w \rho(t)$ where $w$ is a positive or zero constant.

To lowest order Eqs. (\ref{E1}-\ref{E3}) become
\begin{equation}
{\dot{\rho}\over \rho} =-3(w +1){\dot{a_0}\over a_0}=-3 (w +1)H_0,
\end{equation}
\begin{equation}
{2H_0\over a_0}(\dot {\delta a}-H_0 \delta a)={\rho\over 3\gamma\sigma_0^2}-2H_0\dot{\chi}+2H_0^2\chi,
\end{equation}
\begin{equation}
\ddot {\chi} + 3\dot {\chi} H_0={1\over 6\gamma +1}{(1-3w)\rho\over \sigma_0^2},
\end{equation}
and their solutions are
\begin{equation}
\rho(t)=\rho(0)e^{-3(1+w)H_0 t},
\end{equation}
\begin{equation}
\chi={1\over 6\gamma +1}\frac{(1-3w)}{\sigma_0^2}{\rho(0)\over 9H^2_0w}\left({{\rm e}^{-3(1+w)H_0 t}\over 1+w} -{\rm e}^{-3 H_0 t}+{w\over w+1}\right)
\end{equation}
\begin{eqnarray}\label{eq4}
\delta a=&&{(1-3w)\over 6\gamma +1}{\rho(0)a_0(t)\over
9\sigma_0^2H^2_0w}\left [\left(1-{\rm e}^{-3H_0t(1+w)}\right){(4+3w)\over
3(1+w)^2}-{4\over 3}\left(1-{\rm e}^{3H_0t}\right)+{w H_0t\over w +1}\right ]\nonumber\\
&&+{\rho(0)a_0(t)\over 18\gamma(1+w)H_0^2\sigma_0^2}\left(1-{\rm e}^{-3H_0t(1+w)}\right),
\end{eqnarray}
where we have imposed the boundary conditions $\chi(0)=\dot {\chi}(0)=\delta a(0)=0$.

As a consequence of the above, the gravitational constant $G$ acquires a time dependence given by
\begin{equation}
{\dot {G}\over G}\approx -2\dot {\chi}=-{(1-3w)\over 6\gamma +1}{2\rho(t)\over 3H_0w\sigma_0^2}\left(-1+{\rm e}^{+3w H_0t}\right),
\end{equation}
from which:
\begin{equation}
G(t)=G_0 {\rm exp}\left[- {(1-3w)\over 6\gamma +1}{2\rho (t)\over 9H^2_0w \sigma^2_0}
\left({1\over 1+w}-{\rm e}^{+3w H_0t}+{w {\rm e}^{+3(1+w) H_0 t}\over w +1}\right)\right]
\end{equation}

On setting $w=0$ (dust), taking $\rho(t_0)= 2\times 10^{-30}$ g/cm$^3$, $t_0\approx 2\cdot 7\times 10^{17}\sec$  and $H_0\simeq {2\over 3t_0}$ one has 
\begin{equation}
\left.{\dot{G}\over G}\right|_{t=t_0}\approx -10^{-18}{2\gamma\over 8\gamma +1}\sec^{-1},
\end{equation}
and at present $\vert {\dot {G}\over G}\vert_{t=t_0}<2\times 10^{-18}\sec^{-1}$.
Further, again for $w=0$ and $\gamma$ small (actually from solar system
measurements $\gamma\ll 1$), it is sufficient to just consider the last term in Eq. (\ref{eq4}) obtaining:
\begin{equation}
H_T(t_0)=\left.{\ddot a\over a}\right|_{t=t_0}=
 {\dot{a}_0(t_0)\over a_0(t_0)}\left(1+{\delta \dot{a}(t_0)\over \dot {a}_0(t_0)}-{\delta a_0 (t_0)\over a_0 (t_0)}\right)\simeq H_0(1+0.1)
\end{equation}
where $\delta a(t_0)\approx 0.2 a(t_0)$. Finally one obtains for the deceleration parameter
\begin{equation}
q_0(t_0)\approx \left.-{\ddot {a}a\over \dot{a}^2}\right|_{t=t_0}\approx -0.7
\end{equation}

We thus see that, at the present time (just as a quarter of a century ago!),
it is consistent to introduce matter as a
perturbation on a de Sitter universe.


\begin{thebibliography}{00}
\bibitem{Cooper:1982du}
  F.~Cooper and G.~Venturi,
  Phys.\ Rev.\ D {\bf 24} (1981) 3338.
\bibitem{Turchetti:1981zb}
  G.~Turchetti and G.~Venturi,
  Nuovo Cim.\ A {\bf 66} (1981) 221.
\end{thebibliography}
\end{document}